\documentclass[prl,aps,nofootinbib,twocolumn,preprintnumbers,10pt]{revtex4}
\usepackage{amsmath,amssymb,graphicx,epsfig,hyperref,breakurl,color,bm,subfigure,slashed,lipsum,multirow}
\allowdisplaybreaks[4]

\begin{document}

\title{New horizon in particle physics: First observation of CP violation in baryon decays}

\author{Fu-Sheng Yu$^1$\footnote{Corresponding author, Email: yufsh@lzu.edu.cn}, 
Cai-Dian L\"u$^{2,3}$\footnote{Corresponding author, Email: lucd@ihep.ac.cn}}

\affiliation{
$^1$Frontiers Science Center for Rare Isotopes, and School of Nuclear Science and Technology, Lanzhou University, Lanzhou 730000,  China \\
$^2$Institute of High Energy Physics, CAS, P.O. Box 918(4) Beijing 100049, China\\
$^3$School of Physics, University of Chinese Academy of Sciences, Beijing 100049, China}
\pacs{xxx}
\maketitle

A well-known problem in particle physics and cosmology is the observed matter-antimatter asymmetry in the Universe. 
According to the Big Bang theory, equal amounts of matter and antimatter should have been produced in the early universe. 
However, astrophysical observations show that the visible universe today is overwhelmingly dominated by matter.
This asymmetry can only be explained if three key conditions are met: baryon number violation, charge (C) and charge-parity (CP) symmetry violation, and out of equilibrium \cite{Sakharov:1967dj}. 
The CP violation (CPV) in the Standard Model (SM) of particle physics, originating from the mixing of three generations of quarks \cite{Kobayashi:1973fv}, is far too small to account for the observed matter-antimatter asymmetry. 
This discrepancy strongly suggests the existence of new sources of CPV beyond the SM.
Despite 60 years of study since CPV was first observed in 1964 \cite{Christenson:1964fg}, all confirmed cases occur in meson systems ($K$, $B$ and $D$ mesons) \cite{ParticleDataGroup:2024cfk}, with no CPV established in any baryon system.
Given that the visible matter of the universe is predominantly composed of baryons, it is very important to explore CPV in baryon systems.

Baryons cannot undergo mixing and exhibit only direct CPV in their decays, as a result of baryon number conservation. 
Experimental efforts to investigate baryon CPV have been extensively pursued by the BESIII, Belle and LHCb collaborations in recent years \cite{BESIII:2021ypr,Belle:2022uod,LHCb:2016yco,LHCb:2024iis}. 
%CPV in the baryon systems has not yet been well established experimentally until recent days. 
On the other hand,
theoretical predictions on baryon CPV are particularly challenging, as baryons contain one more quark than mesons, making their dynamics significantly complicated. 

A breakthrough in baryon CPV occurred very recently. The LHCb Collaboration reported the first observation of baryon CPV in $\Lambda_b^0\to pK^-\pi^+\pi^-$ decay \cite{LHCb:observation}. 
The total direct CPV is 
\begin{equation}
{A_{CP}(\Lambda_b^0\to pK^-\pi^+\pi^-)=(2.45\pm0.46\pm0.10)\%,}
\end{equation}
which differs from zero by 5.2 standard deviations. 
In particular, the measured CPV in the resonant region of $m_{p\pi^+\pi^-}<2.7$ GeV/$c^2$ is 
\begin{equation}\label{eq:Acp}
    A_{CP}(\Lambda_b^0\to R(p\pi^+\pi^-)K^-)=(5.4\pm0.9\pm0.1)\%,
\end{equation}
with a significance of 6.0 standard deviations.
{This observation is a milestone and a new horizon in particle physics. }

It is interesting that this observation appears in four-body decays.  
The first observations of CP violations in strange, bottom, and charmed meson systems are all in two-body decay processes, such as $K_L^0\to \pi^+\pi^-$  \cite{Christenson:1964fg}, $B^0\to J/\Psi K_S^0$ \cite{Belle:2001zzw,BaBar:2001pki}, and $D^0\to K^+K^-, \pi^+\pi^-$ \cite{LHCb:2019hro}. 
In contrast, the measured CP violations of $\Lambda_b^0\to p\pi^-$ and $pK^-$ are consistent with zero, with uncertainties below $1\%$ \cite{LHCb:2024iis}.
The LHCb Collaboration turns to the study of multi-body $\Lambda_b^0$ decays, working together with theorists.
The decay $\Lambda_b^0\to pK^-\pi^+\pi^-$ has the largest data sample among charmless $b$-baryon decays, with approximately 80000 signal events, resulting in a total CPV uncertainty of just $0.5\%$  \cite{LHCb:observation}. 
The LHCb Collaboration divides the full dataset into four different final-state phase spaces, in order to investigate the resonance contributions to the global CPV.  
The measured CP asymmetries in the four regions have uncertainties of approximately $1\%$, with significant CPV observed in the low invariant mass region of $p\pi^+\pi^-$ as shown in Eq. (\ref{eq:Acp}).  

{The first observation of baryon CPV by LHCb is consistent with the SM expectation.}
It is very intriguing that the measured result for $\Lambda_b^0\to R(p\pi^+\pi^-)K^-$ aligns with the prediction of {a CPV dynamics} involving $N\pi$ rescatterings \cite{Wang:2024oyi}.
The existence of numerous excited nucleon states, $N^*$, which have large uncertainties in their masses and widths, poses significant challenges to distinguish between the states experimentally and to perform theoretical analyses and predictions.
In 2024, {a new theoretical method of CPV dynamics} was proposed, using the scattering data of $N\pi\to p\pi$ or $p\pi\pi$ to circumvent and overcome the problem of large uncertainties of $N^*$ states \cite{Wang:2024oyi}. 
Based on the final-state interaction mechanism,
%which factorizes the short-distance contribution from the long-distance dynamics of final-state interactions.
%$\Lambda_b^0$ decays firstly into a pair of $N\pi$ and an emitted $K^-$ through the short-distance weak interaction in a very short time. 
%Subsequently, the $N\pi$ system undergoes scattering into $p\pi^+\pi^-$ via long-distance dynamics. 
the decay amplitudes can be reliably expressed using a factorization formula for the short-distance contribution and the $N\pi\to p\pi^+\pi^-$ rescattering amplitudes directly obtained from the experimental data for the long-distance dynamics. 
CPV can arise from the weak phases associated with tree-type and loop-type operators, and the strong phases from the amplitudes of $N\pi\to p\pi^+\pi^-$ rescatterings or from the quark loops in the effective Wilson coefficients \cite{Ali:1998eb}. 
The predicted CPV for $\Lambda_b^0\to R(p\pi^+\pi^-)K^-$ is (5.6--5.9)\% \cite{Wang:2024oyi}, which is very consistent with the LHCb measurement in Eq. (\ref{eq:Acp}).
This result demonstrates that the CPV dynamics using $N\pi$ scattering data is suitable for multi-body decays.
Furthermore, this mechanism predicts that the CPV in certain regions of the Dalitz plots could exceed $10\%$, offering promising avenues for future experimental exploration. 

{The observation of baryon CPV plays a crucial role in the search for new physics and in understanding the matter-antimatter asymmetry in the universe.}
The two sources of strong phases are well understood in theory. The quark loops can be perturbatively calculated, while the amplitudes in the $N\pi$ scatterings can be model-independently obtained from the experimental data.
The weak phases are constrained by the comparison between the experimental measurements and the theoretical predictions. 
%The theoretical uncertainties for direct CPV in $\Lambda_b^0\to R(p\pi^+\pi^-)K^-$ are significantly smaller compared to those in meson decays, which usually suffer large ambiguities of strong phases. 
%Therefore, the observation of CPV in $\Lambda_b^0\to R(p\pi^+\pi^-)K^-$ can constrain the parameters of new physics models. 
%opens a new window to explore potential new physics effect and provides valuable insights into the matter-antimatter asymmetry in the universe. 
In this case, new-physics models of baryogenesis can be studied. 

{The observation of CPV in $\Lambda_b$ decays also opens a new frontier in particle physics, particularly in advancing our understanding of strong interaction.} % even though it is consistent with the SM expectation.}   
It is well known that the observation of gravitational waves not only confirms General Relativity but also opens a new era of multi-messenger cosmology. 
Baryons are different from mesons.
Heavy-flavor baryon decays may change the hierarchy of heavy-quark expansions, compared to mesons \cite{Wang:2011uv}.  
%More is different.
%Since the first direct CPV observation in $B\to \pi\pi$ or $K\pi$ decays exceed $10\%$ \cite{ParticleDataGroup:2024cfk}, people expect the similar processes  of $\Lambda_b^0\to p\pi^-$ and $pK^-$ decays with only one more spectator quark, have the same order of CPV. However, no CPV have  been observed here  with uncertainties smaller than $1\%$ after vast efforts \cite{LHCb:2024iis}. 
Besides, with non-zero spins, baryon decays have amplitudes corresponding to different angular momentum configurations.
The puzzle why CPV in $\Lambda_b^0\to p\pi^-$ and $pK^-$ is so small, has recently been addressed in theory through the destruction of CPV induced by the amplitudes of different angular momenta ($L$) \cite{Han:2024kgz,Duan:2024zjv}. 
The CPV of the amplitudes of $L=0$ and $1$ can exceed $10\%$, as large as those of $B$-meson decays, but opposite in sign in the loop-type diagrams of $L=0$ and $1$.   
This represents a new mechanism and a novel phenomenon that does not occur in the corresponding $B$ meson decays, highlighting the unique dynamics of baryon CPV.

In summary, the first observation of baryon CPV by LHCb \cite{LHCb:observation} marks a significant milestone and opens a new horizon in particle physics. 
It offers fresh opportunities to explore new physics, deepen our understanding of the matter-antimatter asymmetry in the universe, and investigate both perturbative and non-perturbative aspects of QCD.
With more data expected from LHCb Run 3 and Run 4 in the near future \cite{Chen:2021ftn}, it is likely that CPV in baryon systems will be observed in more decay processes including bottom and charm baryon decays \cite{Hsiao:2017tif,He:2024pxh},
and through a variety of observables such as angular distributions \cite{Zhang:2021fdd,Wang:2024qff,Wang:2022fih}, providing a richer and more comprehensive understanding of CPV dynamics.

\textit{Acknowledgement}\textemdash  
The authors declare that they have no conflict of interest. 
The work is supported in part by the National Key Research and Development Program of China
(2023YFA1606000) and by the National Natural Science Foundation of China under Grant No.~12275277, 12335003 and 12435004, the Fundamental Research Funds for the Central Universities under No. lzujbky-2023-stlt01 and lzujbky-2024-oy02.

%\nolinenumbers

\begin{thebibliography}{50}
%\cite{Sakharov:1967dj}
\bibitem{Sakharov:1967dj}
Sakharov AD,
``Violation of CP invariance, C asymmetry, and baryon asymmetry of the universe,''
Pisma Zh Eksp Teor Fiz \textbf{5} (1967), 32-35.
%doi:10.1070/PU1991v034n05ABEH002497
%5167 citations counted in INSPIRE as of 24 Mar 2025


%\cite{Kobayashi:1973fv}
\bibitem{Kobayashi:1973fv}
Kobayashi M, Maskawa T,
``CP Violation in the renormalizable theory of weak interaction,''
Prog Theor Phys \textbf{49} (1973), 652-657.
%doi:10.1143/PTP.49.652
%12262 citations counted in INSPIRE as of 24 Mar 2025

%\cite{Christenson:1964fg}
\bibitem{Christenson:1964fg}
Christenson JH, Cronin JW, Fitch VL, \textit{et al.},
``Evidence for the $2\pi$ Decay of the $K_2^0$ Meson,''
Phys Rev Lett \textbf{13} (1964), 138-140.
%doi:10.1103/PhysRevLett.13.138
%4230 citations counted in INSPIRE as of 30 Mar 2025

%\cite{ParticleDataGroup:2024cfk}
\bibitem{ParticleDataGroup:2024cfk}
Navas S \textit{et al.} [Particle Data Group],
``Review of particle physics,''
Phys Rev D \textbf{110} (2024), 030001.
%doi:10.1103/PhysRevD.110.030001
%59 citations counted in INSPIRE as of 21 Aug 2024

%\cite{BESIII:2021ypr}
\bibitem{BESIII:2021ypr}
Ablikim M \textit{et al.} [BESIII],
``Probing CP symmetry and weak phases with entangled double-strange baryons,''
Nature \textbf{606} (2022), 64-69.
%doi:10.1038/s41586-022-04624-1
[arXiv:2105.11155 [hep-ex]].
%73 citations counted in INSPIRE as of 20 Dec 2024

%\cite{BESIII:2018cnd}
%\bibitem{BESIII:2018cnd}
%Ablikim M \textit{et al.} [BESIII],
%``Polarization and entanglement in baryon-antibaryon pair production in electron-positron annihilation,''
%Nature Phys \textbf{15} (2019), 631-634
%doi:10.1038/s41567-019-0494-8
%[arXiv:1808.08917 [hep-ex]].
%196 citations counted in INSPIRE as of 17 Dec 2024

%\cite{Belle:2022uod}
\bibitem{Belle:2022uod}
Li LK \textit{et al.} [Belle],
``Search for CP violation and measurement of branching fractions and decay asymmetry parameters for $\Lambda_c^+\to \Lambda h^+$ and $\Lambda_c^+\to \Sigma^0 h^+$ ($h=K,\pi$),''
Sci Bull \textbf{68} (2023), 583-592.
%doi:10.1016/j.scib.2023.02.017
[arXiv:2208.08695 [hep-ex]].
%39 citations counted in INSPIRE as of 30 Mar 2025

%\cite{HyperCP:2004zvh}
%\bibitem{HyperCP:2004zvh}
%T.~Holmstrom \textit{et al.} [HyperCP],
%``Search for CP violation in charged-Xi and Lambda hyperon decays,''
%Phys. Rev. Lett. \textbf{93} (2004), 262001
%doi:10.1103/PhysRevLett.93.262001
%[arXiv:hep-ex/0412038 [hep-ex]].
%48 citations counted in INSPIRE as of 30 Mar 2025

%\cite{LHCb:2017hwf}
%\bibitem{LHCb:2017hwf}
%Aaij R \textit{et al.} [LHCb],
%``A measurement of the $CP$ asymmetry difference in $\varLambda_{c}^{+} \to pK^{-}K^{+}$ and $p\pi^{-}\pi^{+}$ decays,''
%J High Energy Phys \textbf{03} (2018), 182
%doi:10.1007/JHEP03(2018)182
%[arXiv:1712.07051 [hep-ex]].
%52 citations counted in INSPIRE as of 19 Dec 2024

%\cite{LHCb:2016yco}
\bibitem{LHCb:2016yco}
Aaij R \textit{et al.} [LHCb],
``Measurement of matter-antimatter differences in beauty baryon decays,''
Nature Phys \textbf{13} (2017), 391-396.
%doi:10.1038/nphys4021
[arXiv:1609.05216 [hep-ex]].
%136 citations counted in INSPIRE as of 19 Dec 2024

%\cite{LHCb:2018fly}
%\bibitem{LHCb:2018fly}
%Aaij R \textit{et al.} [LHCb],
%``Search for $C\!P$ violation in $\Lambda^0_b \to p K^-$ and $\Lambda^0_b \to p \pi^-$ decays,''
%Phys Lett B \textbf{787} (2018), 124-133
%doi:10.1016/j.physletb.2018.10.039
%[arXiv:1807.06544 [hep-ex]].
%35 citations counted in INSPIRE as of 19 Dec 2024

%\cite{LHCb:2018fpt}
%\bibitem{LHCb:2018fpt}
%Aaij R \textit{et al.} [LHCb],
%``Search for CP violation using triple product asymmetries in $\Lambda^{0}_{b}\to pK^{-}\pi^{+}\pi^{-}$, $\Lambda^{0}_{b}\to pK^{-}K^{+}K^{-}$ and $\Xi^{0}_{b}\to pK^{-}K^{-}\pi^{+}$ decays,''
%J High Energy Phys \textbf{08} (2018), 039
%doi:10.1007/JHEP08(2018)039
%[arXiv:1805.03941 [hep-ex]].
%44 citations counted in INSPIRE as of 19 Dec 2024

%\cite{LHCb:2019oke}
%\bibitem{LHCb:2019oke}
%Aaij R \textit{et al.} [LHCb],
%``Search for $CP$ violation and observation of $P$ violation in $\Lambda_b^0 \to p \pi^- \pi^+ \pi^-$ decays,''
%Phys Rev D \textbf{102} (2020), 051101
%doi:10.1103/PhysRevD.102.051101
%[arXiv:1912.10741 [hep-ex]].
%42 citations counted in INSPIRE as of 19 Dec 2024

%\cite{LHCb:2019jyj}
%\bibitem{LHCb:2019jyj}
%Aaij R \textit{et al.} [LHCb],
%``Measurements of $CP$ asymmetries in charmless four-body $\Lambda_b^0$ and $\Xi_b^0$ decays,''
%Eur Phys J C \textbf{79} (2019), 745
%doi:10.1140/epjc/s10052-019-7218-1
%[arXiv:1903.06792 [hep-ex]].
%28 citations counted in INSPIRE as of 19 Dec 2024

%\cite{LHCb:2024iis}
\bibitem{LHCb:2024iis}
Aaij R \textit{et al.} [LHCb],
``Measurement of $CP$ asymmetries in $\Lambda_b^0\to ph^{-}$ decays,''
Phys. Rev. D \textbf{111} (2025), 092004.
%doi:10.1103/PhysRevD.111.092004
[arXiv:2412.13958 [hep-ex]].
%8 citations counted in INSPIRE as of 02 Jul 2025

%\cite{LHCb:observation}
\bibitem{LHCb:observation}
Aaij R \textit{et al.} [LHCb], 
``Observation of charge-parity symmetry breaking in baryon decays,''
[arXiv:2503.16954 [hep-ex]].

%\cite{BaBar:2001pki}
\bibitem{BaBar:2001pki}
Aubert B \textit{et al.} [BaBar],
``Observation of CP violation in the $B^0$ meson system,''
Phys Rev Lett \textbf{87} (2001), 091801.
%doi:10.1103/PhysRevLett.87.091801
[arXiv:hep-ex/0107013 [hep-ex]].
%1145 citations counted in INSPIRE as of 31 Mar 2025

%\cite{Belle:2001zzw}
\bibitem{Belle:2001zzw}
Abe K \textit{et al.} [Belle],
``Observation of large CP violation in the neutral $B$ meson system,''
Phys Rev Lett \textbf{87} (2001), 091802.
%doi:10.1103/PhysRevLett.87.091802
[arXiv:hep-ex/0107061 [hep-ex]].
%1273 citations counted in INSPIRE as of 31 Mar 2025

%\cite{Belle:2004mad}
%\bibitem{Belle:2004mad}
%K.~Abe \textit{et al.} [Belle],
%``Observation of large $CP$ violation and evidence for direct $CP$ violation in $B^0 \to \pi^+\pi^-$ decays,''
%Phys. Rev. Lett. \textbf{93} (2004), 021601
%doi:10.1103/PhysRevLett.93.021601
%[arXiv:hep-ex/0401029 [hep-ex]].
%161 citations counted in INSPIRE as of 31 Mar 2025

%\cite{BaBar:2004gyj}
%\bibitem{BaBar:2004gyj}
%B.~Aubert \textit{et al.} [BaBar],
%``Observation of direct CP violation in $B^0 \to K^+ \pi^-$ decays,''
%Phys. Rev. Lett. \textbf{93} (2004), 131801
%doi:10.1103/PhysRevLett.93.131801
%[arXiv:hep-ex/0407057 [hep-ex]].
%428 citations counted in INSPIRE as of 31 Mar 2025

%\cite{LHCb:2019hro}
\bibitem{LHCb:2019hro}
Aaij R \textit{et al.} [LHCb],
``Observation of CP Violation in Charm Decays,''
Phys Rev Lett \textbf{122} (2019), 211803.
%doi:10.1103/PhysRevLett.122.211803
[arXiv:1903.08726 [hep-ex]].
%478 citations counted in INSPIRE as of 31 Mar 2025

%\cite{Wang:2024oyi}
\bibitem{Wang:2024oyi}
Wang JP and Yu FS,
``CP violation of baryon decays with $N\pi$ rescatterings,''
Chin Phys C \textbf{48} (2024), 101002.
%doi:10.1088/1674-1137/ad75f4
[arXiv:2407.04110 [hep-ph]].
%5 citations counted in INSPIRE as of 24 Mar 2025

%\bibitem{SAID}
%The partial wave amplitudes of $N\pi$ scatterings from the SAID program can be seen in the website of https://gwdac.phys.gwu.edu.

%\cite{Ali:1998eb}
\bibitem{Ali:1998eb}
Ali A, Kramer G, Lu CD,
``Experimental tests of factorization in charmless nonleptonic two-body B decays,''
Phys Rev D \textbf{58} (1998), 094009.
%doi:10.1103/PhysRevD.58.094009
[arXiv:hep-ph/9804363 [hep-ph]].
%632 citations counted in INSPIRE as of 24 Mar 2025


%\cite{Wang:2011uv}
\bibitem{Wang:2011uv}
Wang W,
``Factorization of heavy-to-light baryonic transitions in SCET,''
Phys Lett B \textbf{708} (2012), 119-126.
%doi:10.1016/j.physletb.2012.01.036
[arXiv:1112.0237 [hep-ph]].
%40 citations counted in INSPIRE as of 24 Mar 2025


%\cite{Han:2024kgz}
\bibitem{Han:2024kgz}
Han JJ, Yu JX, Li Y, \textit{et al.},
``Establishing CP violation in $b$-baryon decays,''
Phys. Rev. Lett. \textbf{134} (2025), 221801.
%doi:10.1103/ynnx-f63h
[arXiv:2409.02821 [hep-ph]].
%12 citations counted in INSPIRE as of 24 Mar 2025

%\cite{Duan:2024zjv}
\bibitem{Duan:2024zjv}
Duan ZD, Wang JP, Li RH, \textit{et al.},
``Final-state rescattering mechanism of bottom-baryon decays,''
[arXiv:2412.20458 [hep-ph]].
%0 citations counted in INSPIRE as of 24 Mar 2025

%\cite{Chen:2021ftn}
\bibitem{Chen:2021ftn}
Chen S, Li Y, Qian W, \textit{et al.},
``Heavy flavour physics and CP violation at LHCb: a ten-year review,''
Front. Phys. \textbf{18} (2023), 44601.
%doi:10.1007/s11467-022-1247-1
[arXiv:2111.14360 [hep-ex]].
%34 citations counted in INSPIRE as of 31 Mar 2025

%\cite{Hsiao:2017tif}
\bibitem{Hsiao:2017tif}
Hsiao YK, Yao Y, Geng CQ,
``Charmless two-body anti-triplet $b$-baryon decays,''
Phys. Rev. D \textbf{95} (2017), 093001.
%doi:10.1103/PhysRevD.95.093001
[arXiv:1702.05263 [hep-ph]].
%25 citations counted in INSPIRE as of 31 Mar 2025

%\cite{He:2024pxh}
\bibitem{He:2024pxh}
He XG, Liu CW,
``Large CP violation in charmed baryon decays,''
[arXiv:2404.19166 [hep-ph]].
%9 citations counted in INSPIRE as of 31 Mar 2025

%\cite{Zhang:2021fdd}
\bibitem{Zhang:2021fdd}
Zhang ZH,  Guo XH,
``A novel strategy for searching for CP violations in the baryon sector,''
J High Energy Phys \textbf{07} (2021), 177.
%doi:10.1007/JHEP07(2021)177
[arXiv:2103.11335 [hep-ph]].
%6 citations counted in INSPIRE as of 31 Mar 2025

%\cite{Wang:2022fih}
\bibitem{Wang:2022fih}
Wang JP, Qin Q, Yu FS,
``Complementary $CP$ violation induced by $T$-odd and $T$-even correlations,''
Phys. Rev. D \textbf{111} (2025), L111301.
%doi:10.1103/cdgq-tbww
[arXiv:2211.07332 [hep-ph]].
%15 citations counted in INSPIRE as of 24 Mar 2025

%\cite{Wang:2024qff}
\bibitem{Wang:2024qff}
Wang JP, Qin Q, Yu FS,
``CP violation observables in baryon decays,''
[arXiv:2411.18323 [hep-ph]].
%3 citations counted in INSPIRE as of 24 Mar 2025

\end{thebibliography}
\end{document}